\documentclass[12pt]{article}
\usepackage{amssymb}
\usepackage{epsfig}
\usepackage[english]{babel}
\usepackage{float}
\usepackage{slashed}
\newcommand{\be}{\begin{equation}}
\newcommand{\ee}{\end{equation}}
\newcommand{\bea}{\begin{eqnarray}}
\newcommand{\eea}{\end{eqnarray}}

\def\p{\partial}
\def\pslash{\p\raise.3ex \hbox{\kern-.5em /}}
\def\delslash{\nabla\raise.3ex \hbox{\kern-.7em /}}

\begin{document}

\vskip 5cm

\begin{center}
\Large{ \textbf{Non-Hermitian  ${\cal PT}$-Symmetric   Dirac-Pauli
Hamiltonians    with  Real Energy  Eigenvalues in the Magnetic
Field      }}
\end{center}
\vskip 0.5cm\begin{center} \Large{V.N.Rodionov}
\end{center}
\vskip 0.5cm
\begin{center}
{Plekhanov Russian University, Moscow, Russia,  \em E-mail
vnrodionov@mtu-net.ru}
\end{center}

\begin{center}

\abstract{The modified Dirac-Pauli equations, which is entered by
means of ${\gamma_5}$-mass extension of Hamiltonian operators, are
considered. We also take into account the interaction of fermions
with the intensive homogenous magnetic field focusing attention on
 (g-2) gyromagnetic factor of particles with spin $1/2$. Without
 the use of perturbation theory in the external field the
exact energy spectra are deduced with regard to spin effects of
fermions. We discuss the possible proposals of experimental
measuring of properties of new particles which arising in this
model.}

\end{center}

  {\em PACS    numbers:  02.30.Jr, 03.65.-w, 03.65.Ge,
12.10.-g, 12.20.-m}

\section{Introduction}

 Now it is well-known fact, that the reality of the
spectrum in models with a non-Hermitian Hamiltonian is a
consequence of $\cal PT$-invariance of the theory, i.e. a
combination of spatial and temporary parity of the total
Hamiltonian: $[H,{\cal PT}]\psi =0$. When the $\cal PT$ symmetry
is unbroken, the spectrum of the quantum theory is real. This
surprising results explain the growing interest in this problem
which was initiated by Bender and Boettcher's observation
\cite{ben}. For the past a few years has been studied a lot of new
non-Hermitian $\cal PT$-invariant systems (see, for example
\cite{ft12} - \cite{SPEC2}).

The non-Hermitian ${\cal PT}$-symmetric $\gamma_5$-extension of
the Dirac equation is first studied in \cite{ft12} and further
developed in \cite{Rod1}-\cite{RodKr1}. The purpose of this paper
is the continuation of the studying examples of pseudo-Hermitian
relativistic  Hamiltonians, investigations of which was started by
us earlier.

The quantum field theoretic problem of the motion of charged
fermions in a uniform magnetic field has been solved in the paper
\cite{LJ}. In this paper the  energy spectra of the charged
fermions have been obtained as solutions of the Dirac equation at
the neglect  of the anomalous magnetic moment (AMM) of the fermion
with a uniform magnetic field.

New stage of the research of exact solutions in the external field
was opened by the paper Ternov, Bagrov and Zhukovskii \cite{TBZ}.
In this paper, for the first time was found energy spectrum of
fermions moving in the homogeneous magnetic field and was obtained
wave functions for the case of fermions taking into account the
vacuum supplements to the Bohr magneton. After this exact
solutions of the Dirac equation with the Pauli term for charged
and neutral fermions has been confirmed in \cite{1}-\cite{3}.

The exact results, obtained in \cite{TBZ}-\cite{3}, in particular
have been applied to the analysis of the synchrotron radiation and
the neutron decay rates in very strong magnetic fields (see, for
example \cite{{TKR}}). Note also that intensive magnetic fields
exist near and within a number of space objects. So, the magnetic
fields intensity of the order of $10^{12}\div10^{13}$Gauss
observed near pulsars. Here also may be included the recent
opening of such objects as sources soft repeated gamma-ray burst
and anomalous x-ray pulsars. For them magneto-rotational models
are proposed, and they were named as magnetars. It was showed that
for such objects achievable magnetic fields with intensity up to
$10^{15}$Gauss. It is very important that the share of magnetars
in the General population of neutron stars reaches 10\% \cite{4}.
In this regard, we note that the processes with the participation
of neutrinos  in the presence of such strong magnetic fields can
have a significant influence on the processes which may determine
the evolution of astrophysical objects.

In 1965, a hypothesis was proposed by M.A. Markov \cite{Mar},
according to which the mass spectrum of particles should be
limited by "the Planck mass" $m_{Planck} = 10^{19} GeV $.
 The particles with the limiting mass
\be\label{Markov}m \leq m_{Planck}\ee were named "maximons" by the
author. However, condition (\ref{Markov}) initially was purely
phenomenological and it has seemed until recently that the SM can
be applied adequately up to Plank masses. In the current
situation, however, more and more data have accumulated that are
in favor of the necessity of revising some physical principles. In
particular, this is confirmed by abundant evidence that dark
matter, apparently  exists and absorbs a substantial part of the
energy density in the Universe.

In the late 1970s, a new radical approach  was offered by
V.G.Kadyshevsky \cite{Kad1} (see also \cite{KMRS},\cite{Max}), in
which the Markov idea of the existence of a maximal mass of
particles was accepted as a new fundamental principle of the
quantum field theory (QFT). As it is known, the particle mass $m$
in the SM can possess a value in the interval $0 \leq m < \infty$.
In the new geometrical theory, the condition of the mass spectrum
finiteness is postulated  \be\label{M} m \leq {\cal M},\ee where
the maximal mass parameter {\cal M}, which is called by the
\emph{fundamental mass}, is \emph{a new physical constant}. The
quantity ${\cal M}$ is considered as a curvature radius of a five
dimensional hyperboloid whose surface is a realization of the
curved momentum 4-space, or the anti de Sitter space.  Objects
with a mass larger than ${\cal M}$ cannot be regarded as
elementary particles because no local fields correspond to them.
For a free particle, condition (\ref{M}) automatically holds on
surface of a five dimensional hyperboloid. In the approximation
${\cal M} \gg m$ the anti de Sitter geometry goes over into the
Minkowski geometry in the four dimensional pseudo Euclidean space
("flat limit").

We consider ${\cal PT}$-symmetric Hamiltonians from the standpoint
of the algebraic approach developed in works devoted to studying
of non-Hermitian quantum theory \cite{Rod1}-\cite{RodKr1}.
Hamiltonians under $\gamma_5$-extension are non-Hermitian but
{\cal PT}-symmetric. As has been already noted, $H$ is
non-Hermitian due to the summand with $m_2$ changing the sign at
the Hermitian conjugation of the Hamiltonian $(H^+ \neq H)$. It is
easy to verify that $H$ is also non-invariant at individual
transforms $\cal P $ or  $\cal T$  because the summand with $m_2$
changes the sign under impact of any of these transforms. However,
$H$ is invariant with respect to the joint transform. A similar
model was considered in \cite{ft12}, where the {\cal PT}-symmetric
massive Thirring model was investigated in the (1 + 1) dimension.

The inequality $m_1 \geq m_2$ in this theory which is following
from the condition $m^2={m_1}^2-{m_2}^2$, is the basic requirement
that defines a domain of the unbroken symmetry of the Hamiltonian
under study \cite{ft12}. However, this inequality between $m_1$,
$m_2$ and determination of physical mass of particle $m$ are not a
single mass condition. Therefore we can write the new condition
for the physical mass $m$, which, may be more substantial. Indeed,
using the simple mathematical theorem, we can obtain \cite{RodKr}
\be\label{Alm} m\leq {{m_1}^2}/{2m_2}= M.\ee

 We recall that we are investigating the
issue of the existence of constraints on mass parameters in the
given theory. We suggest that there is a constraint on the
parameter $m$.   In this case, there are reasons to believe that a
relationship exists between $M$ obtained by algebraic
transformation and ${\cal M}$ from the \emph{geometric theory with
limited mass}  \cite{Rod1}-\cite{RodKr}.

If one use the notation of fundamental length then we can write
relationship $l_f = 1/{\cal M}$, where $\hbar = c =1$ , which is
mentioned in the works \cite{Kad1} - \cite{Max}). The
non-Hermitian ${\cal PT}$-symmetric Hamiltonians may be considered
as a kind of a very fruitful environment for the creation of new
physics beyond the SM. The pioneering papers in this field were
accomplished by Miloslav Znojil \cite{M1},\cite{M2} where contains
the development theories with {\cal PT}-symmetric Hamiltonians for
considering the possible existence of the \emph{fundamental
length} in quantum physics. In this connection, investigation of
the results of \cite{M1},\cite{M2}, together with studying of
consequences of the limitations of the spectrum of masses of
elementary particles could shed light on the possibility of
comparing the expressions for the fundamental lengths obtained in
different approaches.

Here we are producing our investigation of non-Hermitian  systems
with $\gamma_5$-mass term extension taking into account AMM of
fermions in external magnetic field. We are studying the spectral
and polarization properties of such systems (Section 2.). The
novelty of developed by us approach is associated with predictions
of new phenomena caused by a number of additional terms of the
non-Hermitian Hamiltonians, which radically changes the picture of
interactions (Section 3.). Most intriguing predictions developed
in our paper is devoted to non-Hermitian mass extension
$m\rightarrow m_1+ m_2\gamma_5$ associated with the appearance in
this the algebraic approach of some new particles (Section 4.). It
is important that previously such particles ("exotic particles")
was observed only in the framework of the geometric approach to
the construction of QFT.


\section{ Modified model for the study of non-Hermitian mass parameters}

Let us now consider the solutions of modified Dirac equations for
free massive particles using the  ${\gamma_5}$-factorization of
the ordinary Klein-Gordon operator. In this case similar to the
Dirac procedure one can represent the Klein-Gordon operator in the
form of a product of two commuting matrix operators:

\be\label{D2} \Big({\partial_\mu}^2 +m^2\Big)=
\Big(i\partial_\mu\gamma^{\mu}-m_1-\gamma_5 m_2 \Big)
\Big(-i\partial_\mu\gamma^{\mu}-m_1+\gamma_5 m_2 \Big), \ee where
 the physical mass of
particles $m$ is expressed through the parameters $m_1$ and $m_2$
\be \label{012} m^2={m_1}^2- {m_2}^2. \ee

For so the function would obeyed to the equations of Klein-Gordon
\be\label{KG} \Big({\partial_\mu}^2
+m^2\Big)\widetilde{\psi}(x,t)=0 \ee one can demand that it also
satisfies  one of equations of the first order \be\label{ModDir}
\Big(i\partial_\mu\gamma^{\mu}-m_1-\gamma_5 m_2
\Big)\widetilde{\psi}(x,t)
=0;\,\,\,\Big(-i\partial_\mu\gamma^{\mu}-m_1+\gamma_5 m_2 \Big)
\widetilde{\psi}(x,t)=0 \ee

Equations (\ref{ModDir}) of course, are less common than
(\ref{KG}), and although every solution of one of the equations
(\ref{ModDir}) satisfies to (\ref{KG}), reverse approval has not
designated. It is also obvious that the Hamiltonians, associated
with the equations (\ref{ModDir}), are non-Hermitian, because in
it the $\gamma_5$-dependent mass components appear ($H\neq
H^{+}$):

  \be\label{H} H =\overrightarrow{\alpha} \textbf{p}+ \beta(m_1
+\gamma_5 m_2)\ee  and \be\label{H+} H^+ =\overrightarrow{\alpha
}\textbf{p}+ \beta(m_1 -\gamma_5 m_2).\ee Here  matrices
$\alpha_i=\gamma_0\cdot\gamma_i$, $\beta=\gamma_0$,
$\gamma_5=-i\gamma_0\gamma_1\gamma_2\gamma_3$.   It is easy to see
from (\ref{012}) that the  mass $m$, appearing in the equation
(\ref{KG}) is real, when the inequality \be \label{e210}
{m_1}^2\geq {m_2}^2.\ee is accomplished.

In this section, we will touch upon also question of describing
the motion of Dirac particles, if their own magnetic moment is
different from the Bohr magneton. As it was shown by Schwinger
\cite{Sc}, that  the Dirac equation of particles in the external
electromagnetic field $A^{ext}$ taking into account the radiative
corrections may be represented in the form \be\label{A}
\left({\cal P}\gamma -m\right)\Psi(x)-\int{\cal
M}(x,y|A^{ext})\Psi(y)dy=0, \ee where ${\cal M}(x,y|A^{ext})$ is
the mass operator of fermion in external  field and ${\cal P_\mu}
= p_\mu -e {A^{ext}}_\mu$ . From equation (\ref{A}) by means of
expansion of the mass operator in series
 according to  $ eA^{ext}$ with precision not over then linear
field terms  one can obtain the modified equation. This equation
preserves the relativistic covariance and consistent with the
phenomenological equation of Pauli obtained in his early papers
\cite{TKR}.

Now let us consider the model of massive fermions with
$\gamma_5$-extension of mass $m\rightarrow m_1+\gamma_5 m_2$
taking into account the interaction of their charges and AMM with
the electromagnetic field $F_{\mu\nu}$:

\be\label{Delta} \left( \gamma^\mu {\cal P}_\mu -
 m_1 -\gamma_5 m_2 -\frac{\Delta\mu}{2}\sigma^{\mu \nu}F_{\mu\nu}\right)\widetilde{\Psi}(x)=0,\ee
where $\Delta\mu = (\mu-\mu_0)= \mu_0(g-2)/2$. Here $\mu$ -
magnetic moment of a fermion, $g$ - fermion gyromagnetic factor,
$\mu_0=|e|/2m$ - the Bohr magneton,
$\sigma^{\mu\nu}=i/2(\gamma^\mu \gamma^\nu-\gamma^\nu
\gamma^\mu)$. Thus phenomenological constant $\Delta\mu $, which
was introduced by Pauli,  is part of the equation and gets the
interpretation with the point of view QFT.

The Hamiltonian form of (\ref{Delta}) in the homogenies magnetic
field is the following \be i\frac{\partial}{\partial t}
\widetilde{\Psi}(r,t)=H_{\Delta \mu}\widetilde{\Psi}(r,t),\ee
where \be\label{Delta1} H_{\Delta\mu} = \vec{\alpha}\vec{{\cal P}}
+ \beta(m_1 + \gamma_5 m_2) +
\Delta\mu\beta(\vec{\sigma}\textbf{H}).\ee Given the quantum
electrodynamic contribution in AMM of an electron with accuracy up
to $e^2$ order we have $\Delta\mu=\frac{\alpha}{2\pi}\mu_0 $,
where $\alpha = e^2 =1/137$ - the fine-structure constant and we
still believe that the potential of an external field satisfies to
the free Maxwell equations.

It should be noted that now the operator projection of the fermion
spin at the direction of  its movement  - $\overrightarrow{
\sigma} \overrightarrow{{\cal P }} $ is not commute with the
Hamiltonian (\ref{Delta1}) and hence it is not the integral of
motion. The operator  $\mu_3$, which is commuting with this
Hamiltonian  is
 operator of polarization in the form of the third
component of the polarization tensor \cite{TKR}
 in the direction of the magnetic field, and \be\label{muH}
          \mu_3=m_1\sigma_3 + \rho_2[\vec{\sigma}\vec{{\cal
          P}}]_3
\ee where matrices $$ \sigma_3= \left(%
\begin{array}{cc}
  I & 0 \\
  0 & -I \\
\end{array}%
\right); \,\,\,\,\,              \rho_2 = \left(%
                                      \begin{array}{cc}
                                        0 & -iI \\
                                     iI & 0 \\
                                         \end{array}%
                                       \right).
$$
Subjecting the wave function $\widetilde{ \psi }$ to requirement
to be eigenfunction of the operator polarization $\mu_3$ and
Hamilton operator (\ref{Delta1}) we can obtain: \be\label{Pi}
\mu_3\widetilde{\psi} = \zeta k\widetilde{\psi}, \,\,\, \mu_3=\left(%
\begin{array}{cccc}
  m_1 & 0 & 0 & {\cal P}_1-i{\cal P}_2 \\
  0 & -m_1 & -{\cal P}_1-i{\cal P}_2 & 0 \\
  0 & -{\cal P}_1+i{\cal P}_2 & m_1 & 0 \\
  {\cal P}_1+i{\cal P}_2 & 0 & 0 & -m_1 \\
\end{array}%
\right), \ee where $\zeta=\pm 1$ are characterized the projection
of fermion spin at the direction of the magnetic field, and
 $$H_{\Delta\mu}\widetilde{\psi}=E\widetilde{\psi},$$
  \be\label{Hmu} H_{\Delta\mu}=\left(%
\begin{array}{cccc}
  m_1+H\Delta\mu & 0 & {\cal P}_3 -m_2& {\cal P}_1-i{\cal P}_2 \\
  0 & m_1-H\Delta\mu & {\cal P}_1+i{\cal P}_2  & -m_2-{\cal P}_3\\
  m_2+{\cal P}_3 & {\cal P}_1-i{\cal P}_2 & -m_1-H\Delta\mu & 0 \\
  {\cal P}_1+i{\cal P}_2 & m_2-{\cal P}_3 & 0 & H\Delta\mu-m_1 \\
\end{array}%
\right). \ee

A feature of the model with $ \gamma_5 $-mass contribution is that
it may contain another any restrictions of mass parameters in
addition to (\ref{e210}). Indeed while that for the physical mass
$m$ one may be constructed by infinite number combinations of $
m_1 $ and $ m_2 $, satisfying to (\ref{012}), however besides it
need take into account the rules of conformity of this parameters
in the Hermitian limit $m_2 \rightarrow 0 $. Without this
assumption the developing of Non-Hermitian models may not be
adequate. With this purpose one can determine an additional mass
scale which will depend on $m_1$, $m_2$ and which would put an
upper bound on the mass spectrum of particles. These
considerations make search in the frame of the stringent
restriction of $m\leq m_1$, the existence of more complicated
non-linear dependence on limiting mass value \be\label{mM} m \leq
M(m_1,m_2).\ee   This expression meets the requirements of the
\emph{principle of conformity:} for all \emph{ordinary fermions },
when $M \rightarrow \infty$ we should obtain ordinary Hermitian
theory. In this sense the principle of conformity identical to the
transition to the "flat limit" in the \emph{geometrical model
}(see page 3.).

 Possibly explicit expression for $M(m_1,m_2)$, may be obtained
from the \emph{simple mathematical theorem:} the arithmetical
average of two non-negative real numbers always is not less than
the geometrical mean of the same numbers.
 Really, we have \cite{RodKr}
$$ \frac{{m}^2+{m_2}^2}{2}\geq\sqrt{m^2\cdot {m_2}^2}$$ and
substitution (\ref{012}), we can get the inequality \be\label{mM1}
m\leq {m_1}^2/2m_2 = M(m_1,m_2).\ee
 Values of $M$ is now defined
by two parameters $m_1,m_2$ and in the Hermitian limit, when $m_2
\rightarrow 0$ the value of the maximal mass $M$ tends to
infinity. It is very important that in this limit the restriction
of mass values completely disappear. In such a way one can
demonstrate a natural transition from the Modified Model to the
SM, which contains any values of the physical mass $m$.

Using (\ref{012}) and expression (\ref{mM1}) we can also obtain
the system of two equations \be\label{sys3}\left\{
\begin{array}{c}
  m={m_1}^2-{m_2}^2 \bigskip\\
  M={{m_1}^2}/{2 m_2} \\
\end{array}\right.
\ee  Thus, the solution of this system  relative to the parameters
$m_1$ and $m_2$ may be represented in the form

\be\label{m11111} {m_1}^{\mp} =\sqrt{2}M
\sqrt{1\mp\sqrt{1-m^2/{M}^2}}; \ee

\be \label{m22222}{m_2}^\mp =  M\left(1\mp \sqrt{1-m^2/{M}^2}
\right). \ee

It is easy to verify that the obtained values of the mass
parameters satisfy the conditions (\ref{012}) and (\ref{e210})
regardless of which the sign will be chosen. Besides  formulas
(\ref{m11111}),(\ref{m22222}) in the case of the upper sign are
agreed with conditions

$$m_2\rightarrow 0$$ and $$m_1\rightarrow m$$ when $$M \rightarrow
\infty,$$
 i.e. when the  Hermitian  limit is exist.

 However, if
one choose a lower sign (i.e. for the ${m_1}^+$ and ${m_2}^+$)
such limit is absent. Thus we can see that the nonlinear scheme of
mass restrictions (see (\ref{mM1})) additionally contains the
solutions satisfying to the some new particles. However this
solutions should be considered only as an indication of the
principal possibility of the existence of such particles. In this
case, as follows from (\ref{m11111}),(\ref{m22222}), for each of
ordinary particle may be exist the new partner, possessing the
same mass and possibly having  a number of another similar
properties.\footnote{As the exotic particles do not agree in the
"flat limit"
 with the ordinary Dirac expressions then one can
assume that in this case we deal with a description of some new
particles, properties of which have not yet been studied. This
fact for the first time has been fixed by V.G.Kadyshevsky in his
early works in the geometric approach to the development of the
quantum field  theory with a fundamental mass" Ref.\cite{Kad1} in
curved de-Sitter momentum space. Besides in
Ref.\cite{KMRS},\cite{Max} it was noted that the most intriguing
prediction of the new approach is the possible existence of exotic
fermions with no analogues in the SM, which may be candidates for
constituents of dark matter .}

\begin{figure}[h]
\vspace{-0.2cm} \centering
\includegraphics[angle=0, scale=0.5]{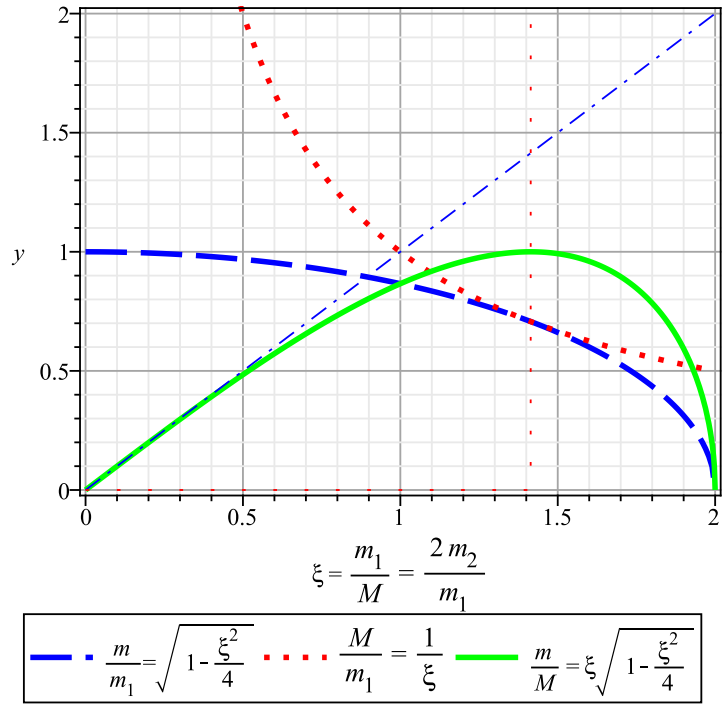}
\caption{Dependence of $m/M,  M/m_1$, and  $m/m_1 $ on the
parameter $\xi=m_1/M=2m_2/m_1$} \vspace{-0.1cm}\label{Fig.1-1}
\end{figure}

Let's consider the  "normalized" parameter of the modified model
with the maximal mass $M$: \be \label{x} \xi=\frac{m_1}{M}=\frac{2
m_2}{m_1}\ee and using (\ref{012}), we can obtain \be\label{y}
\frac{m_2}{M}=\frac{\xi^2}{2}\ee \be
\frac{m}{M}=\xi\sqrt{1-\xi^2/4}\ee

At Fig.1 one can see the dependence of the normalized parameters
$m/m_1$, $M/m_1$ and $m/M$ on the relative parameter
$\xi=m_1/M=2m_2/m_1$. In particular, the maximum value of the
particle mass $ m = M $ is achieved at the ratio of the subsidiary
masses is equal to $m_2 = m_1/ \sqrt{2}$. Till to this value for
each mass of ordinary particles, one can find the parameters $m_1$
and $m_2$, for which a limit transition to regular Dirac theory
exist. Further increasing of $m_2$, leads to the descending branch
of the $m/M$, where the   Dirac limit  is  not exist  and at the
point $m_2=m_1$ the value of $m$ is equal to zero. Thus, it is the
region $m_1 > \sqrt{2}m_2$ ($m_2 > M$ ) corresponds to the
description of the "exotic particles", for which there is not
transition to Hermitian limit. Note, that the equality $m_2=m_1$
corresponds to the case of massless exotic fermions.

\section{ Exact solutions of Dirac-Pauli equations in the intensive
uniform  magnetic field}

Performing calculations  in many ways reminiscent of similar
calculations carried out in the ordinary model in the magnetic
field \cite{TKR}, in a result, for modified Dirac-Pauli equation
one can find \emph{the exact solution for energy spectrum}
\cite{Rod5}:
 \be\label{E61} E(\zeta,p_3,2\gamma
n,H)=\sqrt{{p_3}^2-{m_2}^2+\left[\sqrt{{m_1}^2+2\gamma
n}+\zeta\Delta\mu H \right]^2} \ee and for eigenvalues of the
operator polarization $\mu_3$ we can write in the form \be
k=\sqrt{{m_1}^2 +2\gamma n}. \ee

\begin{figure}[h]
\vspace{-0.2cm} \centering
\includegraphics[angle=0, scale=0.5]{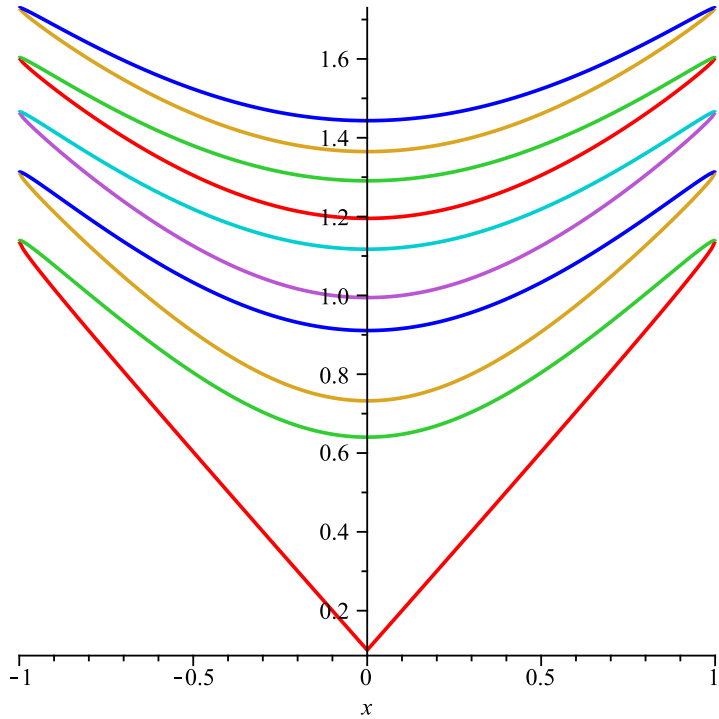}
\caption{Dependence of $E(+1,0, 0.4n, 0.1 ) $ on the parameter
$x=m/M$ for the cases $n=0,1,2,3,4\,\, and\,\,\Delta\mu H = 0.1.$}
\vspace{-0.1cm}\label{Fig.1-p1}
\end{figure}

From (\ref{E61}) it follows that in the field  where  ${\cal PT}$
symmetry is unbroken $m \leq M$, all energy levels are real for
the case of spin orientation along the magnetic field direction
$\zeta =+1$. We can clearly see the dependence of the set of
energy values from the parameter $x=m/M$ on Fig.\ref{Fig.1-p1}.

\begin{figure}[h]
\vspace{-0.2cm} \centering
\includegraphics[angle=0, scale=0.5]{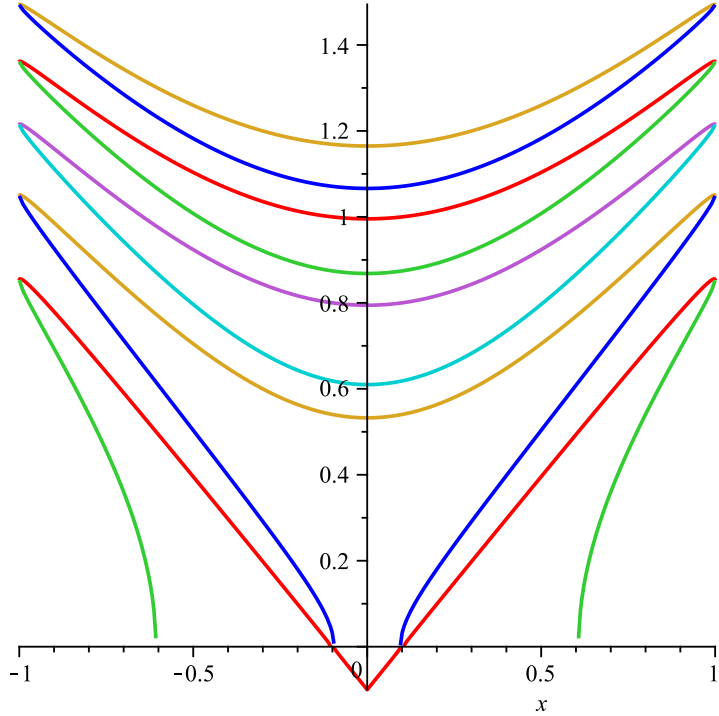}
\caption{Dependence of $E(-1,0, 0.4n, 0.1 )$ on the parameter
$x=m/M$ for the cases $n=0,1,2,3,4\,\, and\,\,\Delta\mu H = 0.1.$}
\vspace{-0.1cm}\label{Fig.2-p1}
\end{figure}

\begin{figure}[h]
\vspace{-0.2cm} \centering
\includegraphics[angle=0, scale=0.5]{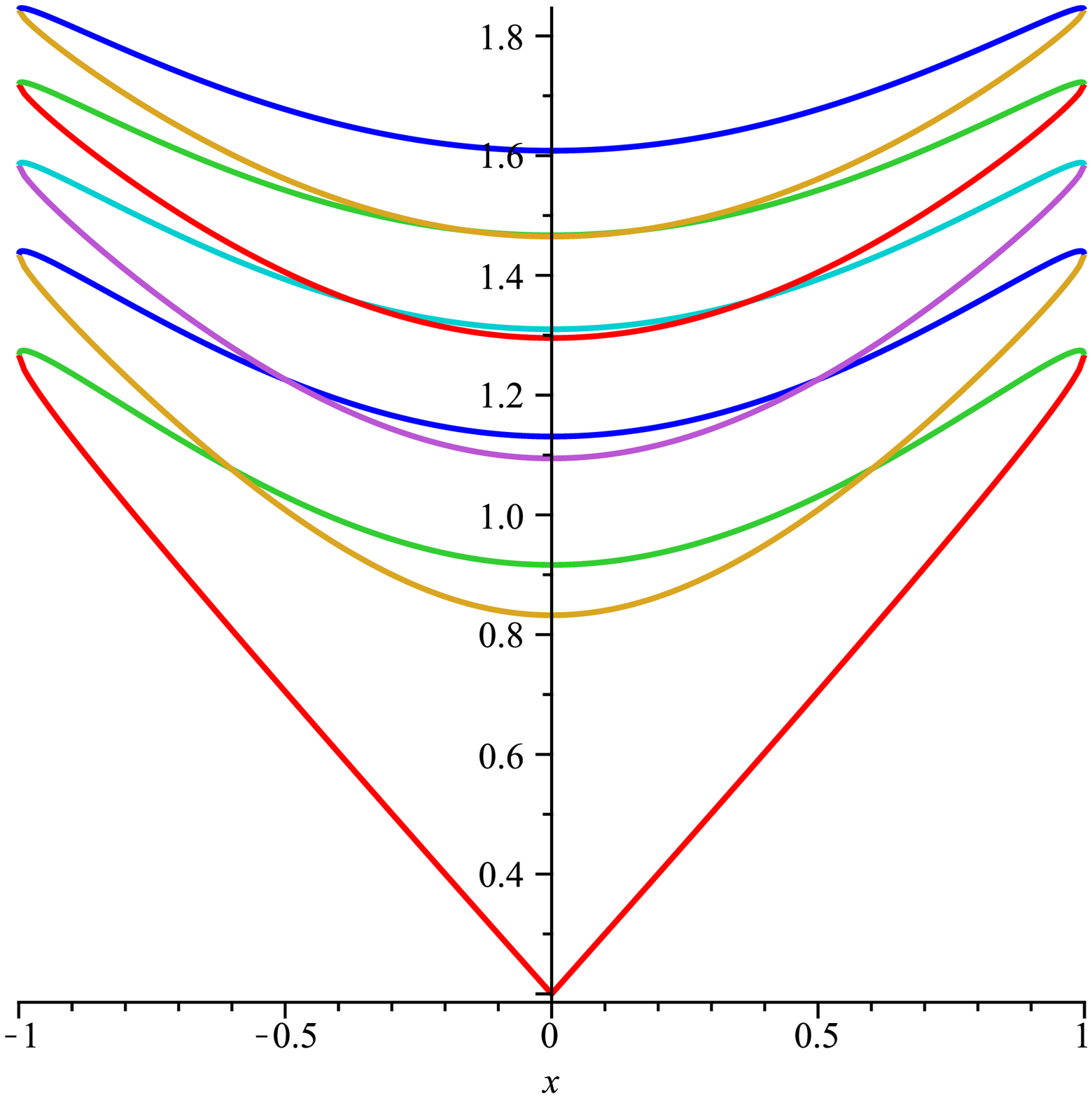}
\caption{Dependence of $E(+1,0, 0.4n, 0.2 ) $ on the parameter
$x=m/M$ for the cases $n=0,1,2,3,4\,\, and\,\,\Delta\mu H = 0.2.$}
\vspace{-0.1cm}\label{Fig.3-p1}\end{figure}

\begin{figure}[h]

\vspace{-0.2cm} \centering
\includegraphics[angle=0, scale=0.5]{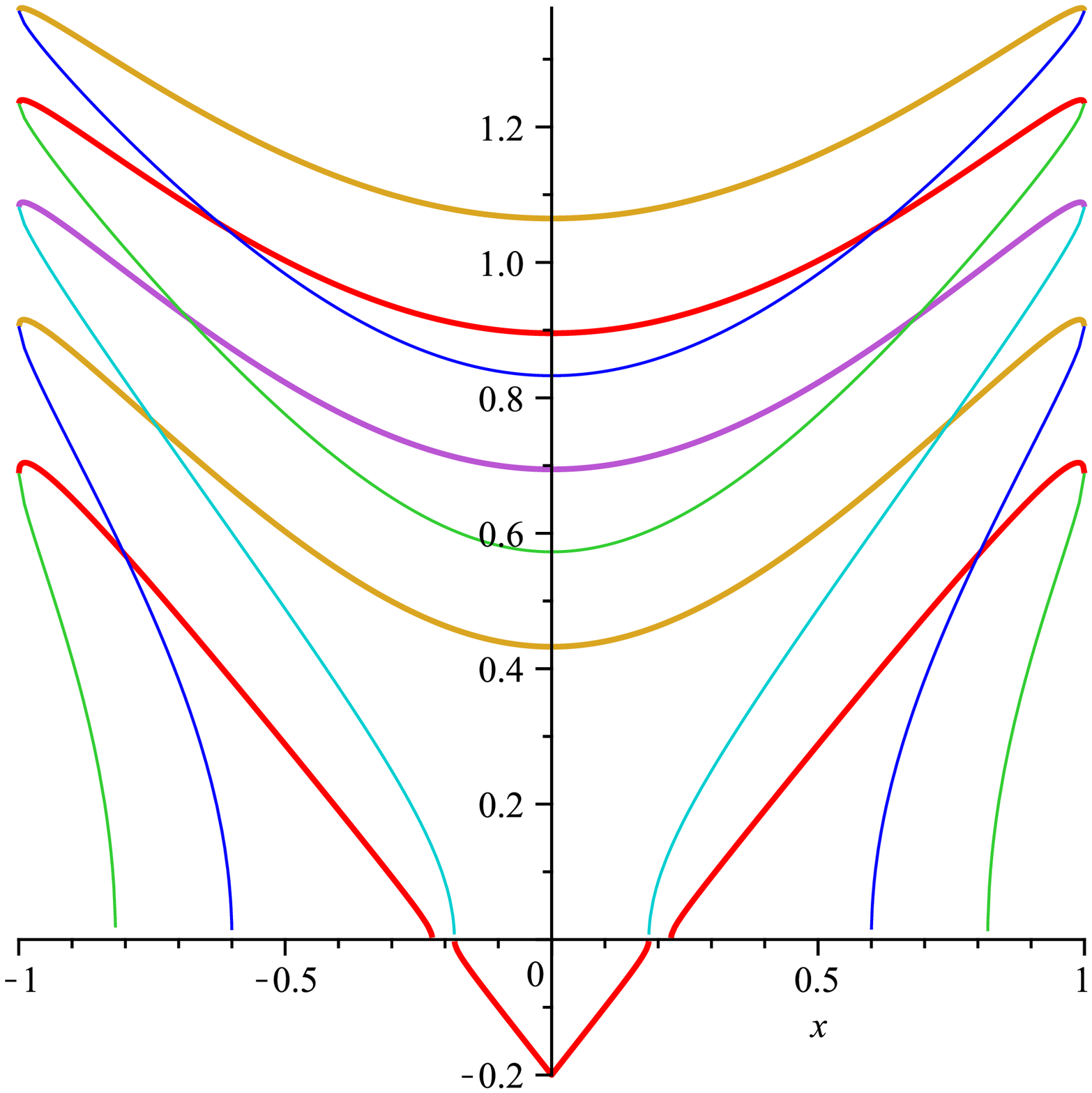}
\caption{Dependence of $E(-1,0, 0.4n, 0.2) $ on the parameter
$x=m/M$ for the cases $n=0,1,2,3,4\,\, and\,\,\Delta\mu H = 0.2.$}
\vspace{-0.1cm}\label{Fig.4-p1}
\end{figure}

However, in the opposite case $\zeta = -1$ we have the  imaginary
part from the ground state of fermion $n=0$ and other low energy
levels, see on Fig.\ref{Fig.2-p1}. For the cases of increasing
parameter $\Delta\mu H = 0.2$ we can watch overlapping of
different levels (see  Fig.\ref{Fig.3-p1} and Fig.\ref{Fig.4-p1}).


It is easy to see that in the case $ \Delta\mu =0$ from
(\ref{E61}) one can obtain the ordinary expression for energy of
charged particle in the magnetic field (\emph{Landau levels}).
Besides it should be emphasized that from the expression
(\ref{E61}), in the Hermitian limit putting $m_2=0$ and $m_1=m$
one can obtain: \be\label{DPA} E(\zeta,p_3,2\gamma
n,H)=\sqrt{{p_3}^2+\left[\sqrt{{m}^2+2\gamma n}+\zeta\Delta\mu H
\right]^2}.
 \ee

Note that in the paper \cite{TBZ} was previously obtained result
analogical to  (\ref{DPA}) by means of using of the Hermitian
Dirac-Pauli approach. Direct comparison of formula (\ref{DPA})
with the  result \cite{TBZ} shows their coincidence. It is easy to
see that the expression (\ref{E61}) contains dependence on
parameters $m_1$ and $m_2$ separately, which are not combined into
\emph{a mass of particles}, that essentially differs from the
examples which were considered early \cite{ft12}-\cite{RodKr}.

 Thus, here the calculation of
interaction AMM of fermions with the magnetic field allow to put
the question about the possibility of experimental studies of the
non-Hermitian effects of $\gamma_5$-extensions of a fermion mass.
Thus, taking into account the expressions (\ref{m11111}) and
(\ref{m22222}) we obtain that the energetic spectrum (\ref{E61})
is expressed through the fermion mass $m$ and the value of the
maximal mass $M$. Thus, taking into account that the interaction
AMM with magnetic field removes the degeneracy on spin variable,
we can obtain the energy of the ground state ($\zeta=-1$) in the
form
 \be\label{E1}
E(-1,0,0,H,x)=m\sqrt{-\left({\frac{1\mp\sqrt{1-x^2}}{x}}\right)^2+
\left(\frac{\sqrt{2}\sqrt{1\mp\sqrt{1-x^2}}}{x}-\frac{\Delta\mu
H}{m} \right)^2}, \ee where $x=m/M$ and the upper sign corresponds
to the ordinary particle and the lower sign defines their "exotic"
partners.

Through decomposition of functions ${}^{-}m_1$ and ${}^{-}m_2$ we
can obtain \be\label{sys}{}^{-}m_1/m =\left\{
\begin{array}{c}
  1+\frac{x^2}{8}+\frac{7 x^4}{128},\,x\ll 1 \bigskip\\
  \frac{\sqrt{2}}{x}\qquad \qquad x\rightarrow 1 \\
\end{array}\right.\,\,\,\,\,
{}^{-}m_2/m =\left\{
\begin{array}{c}
   \frac{x}{2}+\frac{x^3}{8}+\frac{x^5}{16},\,\,\,\,\,\,\,\,x\ll 1\bigskip\\
  \frac{1}{x}\qquad\qquad\qquad x\rightarrow 1 \\
\end{array}\right.
\ee Similarly, for ${}^{+}m_1$ and ${}^{+}m_2$ we have

\be\label{sys1}{}^{+}m_1/m =\left\{
\begin{array}{c}
  \frac{2}{x}-\frac{x}{4}-\frac{5 x^3}{64},\,x\ll 1\bigskip \\
  \frac{\sqrt{2}}{x}\qquad \qquad x\rightarrow 1 \\
\end{array}\right.\,\,\,\,\,
{}^{+}m_2/m =\left\{
\begin{array}{c}
   \frac{2}{x}-\frac{x}{2}-\frac{x^3}{8},\,x\ll 1\bigskip\\
  \frac{1}{x}\qquad\qquad\,\, x\rightarrow 1 \\
\end{array}\right.
\ee

Thus, it is shown that the main progress, is obtained by us in the
algebraic way of the construction of the fermion model with
$\gamma_5$-mass term is consists of describing of the new
energetic scale, which is defined by the parameter
$M={m_1}^2/2m_2$. This value on the scale of the masses is a point
of transition from the ordinary particles $m_2 < M$ to exotic $m_2
> M $. Furthermore, description of the exotic fermions in the
algebraic approach are turned out essentially the same as in the
model with a maximal mass, which was investigated by
V.G.Kadyshevsky with colleagues on the basis of geometrical
approach \cite{Kad1}-\cite{Max}.

It should be noted that the formula (\ref{E61})  is a valid not
only for charged fermions, but and for the neutral particles
possessing AMM. In this case one must simply replace the value of
quantized transverse momentum of a charged particle in a magnetic
field on the ordinary value $2\gamma n\rightarrow
{p_1}^2+{p_2}^2$.
 Thus, for the case of ultra cold polarized ordinary electronic
 neutrino with precision not over then linear
field terms   we can write

\be\label{E34} E(-1,0,0,H,M \rightarrow \infty)= m_{\nu_e}
\sqrt{1-\frac{\mu_{\nu_e}}{\mu_0}\frac{
 H}{ H_c}}.
 \ee
However, in the case of exotic electronic
 neutrino we have
 \be\label{E3}
E(-1,0,0,H,m_{\nu_e}/M)= m_{\nu_e}
\sqrt{1-\frac{\mu_{\nu_e}}{\mu_0}\frac{2 M
 H}{m_{\nu_e} H_c}}.
 \ee

It is well known \cite{n3},\cite{n33} that in the minimally
extended SM the one-loop radiative correction generates neutrino
magnetic moment which is proportional to the neutrino mass
\be\label{mu1}
  \mu_{\nu_e}=\frac{3}{8\sqrt{2}\pi^2}|e| G_F
  m_{\nu_e}=\left(3\cdot10^{-19}\right)\mu_0\left(\frac{m_{\nu_e}}{1
  eV}\right),
\ee where $ G_F$-Fermi coupling constant and $\mu_0$ is Bohr
magneton.
 However, so far, the most stringent laboratory constraints on the
 neutrino magnetic moment come from elastic
 neutrino-electron scattering experiments:
$ \mu_{\nu_e} <(1.5\cdot 10^{-10})\mu_0$\cite{n1}. Besides the
discussion of problem of measuring the mass of neutrinos (either
active or sterile) show that for an active neutrino model we have
$\sum m_\nu =0.320 eV$, whereas for a sterile neutrino $\sum m_\nu
=0.06 eV$ \cite{n2}.

\section{Conclusions}
One can also estimate the change of the border of region of
unbroken ${\cal P}{\cal T}$-symmetry due to the shift of the
lowest-energy state in the magnetic field. Using formulas
(\ref{E34}) and (\ref{E3}) we obtain correspondingly regions of
undisturbed  ${\cal P}{\cal T}$-symmetry  in the form
\be\label{border} H_{\nu_e (ordinary)}\leq\frac{{\mu_0}}{
\mu_{\nu_e}} H_c ;\ee \be\label{border1} H_{\nu_e
(exotic)}\leq\frac{m_{\nu_e}{\mu_0}}{2 M \mu_{\nu_e}} H_c. \ee

 Indeed let us take  the
following parameters of neutrino: the mass of the electronic
neutrino is  equal to $m_{\nu_e} = 1 eV$ and magnetic moment equal
to (\ref{mu1}). If we assume that the values of mass and magnetic
moment of exotic neutrino identical to parameters of ordinary
neutrinos, we can obtain the estimates of the border area
undisturbed ${\cal P}{\cal T}$ symmetry  for (\ref{border}) in the
form \be\label{E4}
 {H^{cr}}_{\nu_e (ordinary)}  =  \frac{\mu_0}{\mu_{\nu_e}} H_c \sim 10^{32} Gauss.\ee
However in the case (\ref{border1}) the situation may change
radically \be\label{E41}
 {H^{cr}}_{\nu_e(exotic)}  =  \frac{\mu_0}{\mu_{\nu_e}} \frac{m_{\nu_e}}{2 M} H_c \sim 10^4 Gauss.\ee
 In (\ref{E4}) and (\ref{E41}) we used the values of
quantum-electrodynamic constant $H_c=4.41\cdot 10^{13}$Gauss and
the Planck mass $M = m_{Planck}\simeq 10^{19}GeV$.

In the case of (\ref{E4}) one can see that the experimentally
possible field corrections are extremely small, because the
critical values of the magnetic field are fantastic large.  On the
other hand it is obvious that the critical value of magnetic field
(\ref{E41}) is attainable in the sense of ordinary terrestrial
experiments. We do not know if there is an upper limit of spectrum
masses of elementary particles consistent with the Markov's
conjecture \cite{Mar}.  However contemporary precision of
alternative laboratory measurements at low energy in the magnetic
field may in principle allow to achieve the required values of
exotic particles in the near future. Thus, consequences from the
obtained formulas (\ref {E3}) - (\ref {E41}) allow to be convinced
not only in the existence of the Maximal Mass but and in reality
of the so-called \emph {exotic particles}, because this phenomena
are inextricably related.

 {\bf
Acknowledgment:} We are grateful to Prof. V.G.Kadyshevsky for
fruitful and highly useful discussions.

 \end{document}